\documentstyle[multicol,prl,aps,epsf,epsfig]{revtex}

\def \b{\beta}

\def \be{\begin{equation}}
\def \ee{\end{equation}}
\def \ben{\begin{eqnarray}}
\def \een{\end{eqnarray}}

\begin{document}

\title{Primordial braneworld black holes: significant enhancement of
lifetimes through accretion}
 
\author{A. S. Majumdar\footnote{Electronic address: archan@bose.res.in}}
 
\address{S. N. Bose National Centre for Basic Sciences,
Block JD, Sector III, Salt Lake, Kolkata 700098, India}

\maketitle

\begin{abstract}

The Randall-Sundrum (RS-II) braneworld cosmological model with a fraction 
of the total energy density in primordial black holes is considered. 
Due to their 5-d geometry these
black holes undergo modified Hawking evaporation.  
It is shown that during the high energy regime accretion from 
the surrounding radiation bath is dominant compared to evaporation. 
This effect increases the mass of the black holes till  the onset of matter 
(or black hole) domination of the total energy density. Thus black holes with 
even very small initial masses could
survive till several cosmologically interesting eras.

\end{abstract}

Primordial black holes are formed through various mechanisms\cite{carr} in the 
early universe. Depending on their evolution and evaporation time, relic
black holes could have significant cosmological consequences. For
instance, they could be responsible for the generation of the observed 
baryon asymmetry in the universe\cite{majumdar}. Futhermore, depending
on how long the relic black holes survive, these could also serve as
viable candidates for cold dark matter\cite{blais}. The key question in the
cosmology of primordial black holes is the duration of their lifetimes,
vis-a-vis their initial mass spectrum and formation times. In this
article we focus on this issue in the context of the RS-II\cite{randall}
braneworld model.

In the RS-II braneworld scenario matter and radiation are confined to
the 3-brane, whereas gravity propagates also in the bulk. The cosmological
ramifications of this model have received much attention 
recently\cite{langlois}. In this model
 there exists a regime during the early
stages when the expansion rate of the universe is proportional to its 
energy density. However, the standard cosmological evolution is recovered
for times $t>> t_c$, where $t_c$ is related to $l$, the size of the extra 
dimension. The present experimental limit\cite{long} to this size is 
$l \le 0.1{\rm mm}$. Our purpose here is to consider the black holes
which are formed in the early (high energy) phase of this braneworld
model. Such black holes obey a modified Hawking evaporation law due to
a different induced metric on the brane (compared to standard 4d black
holes) and also because of the radiation of a part of the gravitational
energy into the bulk\cite{guedens}. Moreover, accretion from the surrounding
radiation bath could have a significant effect on the evolution of these
black holes\cite{majumdar2,guedens2}. In what follows we describe in 
detail the effects of the interplay of evaporation and accretion on the
lifetime of a population of primordial black holes in the braneworld.

We begin with an early era in the high energy braneworld phase in which 
a certain number density $n_{BH}$ of primordial
black holes with individual mass $M$ exchange energy with the 
surrounding radiation by accretion and evaporation. Let
the fraction of the total energy in black holes at some initial time $t_0$ be
$\b$. Hence, the cosmological evolution will be governed by the Friedmann
equation, the evolution equation for black hole mass, and the equation
for radiation density incorporating the effects of evaporation and accretion.
These three equations (for $t \ll t_c$) are given by
\be
\label{1}
{\dot{a}^2\over a^2} = \Biggl({8\pi t_c\over 3M_4^2}\Biggr)^2\biggl(Mn_{BH} + \rho_R\biggr)^2,
\ee
\be
\label{2}
\dot{M} = -{AM_4^2 \over Mt_c} + {64 M \rho_R t_c\over 3 M_4^2}
\ee
and
\be
\label{3}
{d \over dt}\biggl(\rho_R(t)a^4(t)\biggr) = - \dot{M}(t)n_{BH}(t)a(t)
\ee
where $(A \simeq  3/(16)^3\pi)$ and  $M_4$ ($l_4$)
are the 4d Planck mass (length).

A full numerical integration of the above equations would lead to the
general description of the cosmology we are considering. However, here
we assume the condition of radiation domination, i.e., $\b \ll 1$ 
($\rho_R \gg Mn_{BH}$) to hold. (Note that we have already made another
simplifying assumption by choosing the same initial mass $M_0$ and formation
time $t_0$ for all black holes, instead of considering a realistic mass
spectrum). Under these assumtions Eqs.(1) and (3) simplify to yield
$a \propto t^{1/4}$ and $\rho_R \propto 1/t$ during the high energy
brane phase, i.e., for $t \le t_c$. For later times, the standard
radiation dominated expansion is recovered. It was shown in \cite{majumdar2}
that during the radiation dominated high energy phase, the solution of
the black hole equation (2) is such that the mass of an individual black
hole continues to grow as
\be
\label{4}
{M(t) \over M_0} \simeq \Biggl({t\over t_0}\Biggr)^B
\ee
where $(B \simeq 2/\pi)$.
Due to an imperfect efficiency of accretion, the value of the
number $B$  might get reduced by an $O(1)$
factor\cite{guedens2}. Nevertheless, this result indicates the dominance
of accretion over evaporation in the dynamics of primordial braneworld
black holes.

It can be further shown\cite{majumdar2} that for the condition of radiation 
domination to hold up to $t_c$, one must have
\be
\label{5}
{\b \over 1 -\b} < \Biggl({t_0 \over t_c}\Biggr)^{B + 1/4}
\ee
Beyond $t_c$, the cosmology enters the standard low energy phase. During
this stage the black hole masses continue to grow up to a certain time
$t_t$  after which the surrounding radiation 
gets dilute enough for accretion to become negligible. Henceforth, 
evaporation takes over the dynamics, and
the solution of the black hole equation (2) can be approximated by
\be
\label{6}
M(t) = \Biggl[M^2(t_t) - {2AM_4^2\over t_c}(t - t_t)\Biggr]^{1/2}
\ee
From Eq.(6) the lifetime of black holes is derived\cite{majumdar2} to be
\be
\label{7}
{t_{end} \over t_4} \simeq {4\over A}(2\sqrt{2})^B\Biggl({M_0\over M_4}\Biggr)^{2-B}
{t_c\over t_4}\Biggl({t_t^2 \over t_c t_4}\Biggr)^B
\ee
The exact computation of $t_t$ can only be done by a full numerical simulation
of Eqs.(1--3). In comparison with the the results for primordial black holes 
in standard
cosmology, one finds that the lifetime of braneworld black holes are
significantly longer. Although the modified evaporation law and accretion
are both responsible for this effect, the latter is certainly the more
effective reason. For example, with the inclusion of accretion in a suitable
parameter range, subplanckian ($M_0 < M_4$) black
holes could survive up to the era of nucleosynthesis, and those with
initial mass $M_0 \simeq 10^3{\rm g}$ could survive up to the present 
era\cite{majumdar2}.

To summarize, we have seen that accretion from the surrounding radiation
bath is an important effect for primordial black holes in the high energy 
phase of the RS-II braneworld scenario. For a wide range of values of 
the initial
black hole mass and the size of the extra dimension, accretion indeed
dominates over evaporation. Significant enhancement of
the black hole lifetime occurs. Consequently, several astrophysical
constraints on primordial black holes are impacted\cite{guedens2}. The 
multifarous consequences of long-lived black holes in cosmology need to be 
investigated in more details.

\end{document}